\documentclass{emulateapj}

\slugcomment{Accepted in ApJ Part I, 2004, October 6}

\shorttitle{OH rotational lines toward NGC 253 and NGC 1068}
\shortauthors{Goicoechea et al.}

\begin{document}

\title{OH rotational lines  as a diagnostic of the warm neutral gas in
galaxies\altaffilmark{1}}

\author {Javier R. Goicoechea\altaffilmark{2}, Jesus Mart\'{\i}n--Pintado
and Jos\'e Cernicharo}

\affil{Departamento de Astrofisica Molecular e Infrarroja, Instituto de 
Estructura de la Materia, CSIC, Serrano 121, 28006, Madrid, Spain}

\email{javier@damir.iem.csic.es}

\altaffiltext{1}{Based on observations with ISO, 
an ESA project with instruments funded by ESA Member States 
(especially the PI countries: France, Germany, the Netherlands 
and the United Kingdom) and with participation of ISAS and NASA.} 

\altaffiltext{2}{On leave to the Laboratoire d'\'{E}tude du Rayonnement et de la 
Mati\`ere, 
UMR 8112, CNRS, \'{E}cole Normale Sup\'{e}rieure et Observatoire de Paris,
24 rue Lhomond,  75231 Paris Cedex 05, France}

\begin{abstract}
We present \textit{Infrared Space Observatory} (ISO) observations of 
several OH, CH and H$_2$O rotational lines toward the bright infrared 
galaxies NGC~253 and NGC~1068. As found in the Galactic clouds in Sgr~B2 and
Orion, the extragalactic far--IR OH lines change from 
absorption to emission depending on the  physical conditions 
and distribution of gas and dust along the line of sight. 
As a result, most of the OH rotational lines that appear in absorption
toward NGC~253 are observed in emission toward NGC~1068.
We  show that  the far--IR spectrum of OH can be used as a powerful diagnostic
to derive the physical conditions of extragalactic neutral gas. In particular,
we find  that a warm 
(T$_k$$\sim$150~K, $n$(H$_2$)$\lesssim$5$\times$10$^4$~cm$^{-3}$) component
of molecular gas with an OH abundance of $\simeq$10$^{-7}$ from the inner
$\lesssim$15$''$ can qualitatively reproduce the OH lines toward NGC~253. 
Similar temperatures but higher densities ($\sim$5$\times$10$^5$~cm$^{-3}$) are 
required to explain the OH emission in NGC~1068.
\end{abstract}


\section{Introduction}

Galactic nuclei play a key role for our understanding of galactic evolution.
The use of molecular diagnostics to extract their physical conditions helps
to understand  the nature of the prevailing scenarios, e.g., starbursts vs.
active galactic nuclei (AGN).
Space observations at mid-- and far--IR wavelengths offer an 
unique opportunity to observe the pure rotational lines of 
light hydrides, the simplest chemical building blocks.
In particular, H$_2$ observations have revealed the presence of a
considerable amount of \textit{``warm''} 
(excitation temperatures, T$_{ex}$, around 100-200~K) neutral gas
in the Galactic Center (GC; Rodr\'{\i}guez--Fernandez et al. 2001) and also 
in galaxies (Rigopoulou et al. 2002).

\textit{ISO} has given the opportunity to observe polar species 
that are  sensitive to a broad range of densities and kinetic temperatures.
The far--IR spectrum, and thus the molecular content,
of molecular clouds such as Sgr~B2 in the GC (Goicoechea et al. 2004)
is surprisingly similar to that observed in ultra-luminous IR galaxies
such as Arp~220 (Gonz\'alez--Alfonso et al. 2004).  
One of the  species with the largest abundance is the OH radical.

Bright IR galaxies containing OH usually show megamaser emission at
18~cm. These lines arise from the hyperfine splitting of the OH 
$^2\Pi_{3/2}$ $J$=3/2 ground rotational state.
Among the possible excitation mechanisms of these masers; near-- and far--IR
pumping, collisions and UV radiation, the far--IR excitation seems to be the
dominant mechanism inverting the population of the hyperfine levels 
(e.g. Skinner et al. 1997).
Furthermore, the relative population of the OH  levels 
is governed by the far--IR pure
rotational lines and detailed treatment of the rotational excitation is required 
to analyze the OH radio lines at 18~cm but also the 
$^2\Pi_{3/2}$ $J$=5/2 and $^2\Pi_{1/2}$ $J$=1/2 lines
at 5 and 6~cm (Randell et al. 1995).

The OH ground--state rotational line at $\sim$119~$\mu$m
was first detected in absorption toward Sgr~B2 (Storey et al. 1981),
while several excited rotational lines have been analyzed
by Goicoechea \& Cernicharo (2002).
A detailed analysis of the far--IR  spectrum of OH has been also carried
out in the Orion region (Watson et al. 1985; Vicuso et al.
1985; Melnick et al. 1987, 1990; Betz \& Boreiko 1989). 
The observed lines appear in emission and some of them even show more
complex P--Cygni profiles. Different excitation conditions determine
the distinct OH spectral signatures observed toward Orion and Sgr~B2.

These studies have shown that OH can be a powerful prove of the warm
neutral gas in molecular clouds. Scaled to extragalactic studies, OH can
thus provide significant information about star forming regions
and the heating processes in the inner arcsecs of a galaxy, in addition
to the understanding of the megamaser excitation mechanism.
Moreover, OH is a key molecule in the gas--phase chemistry networks.
The presence of OH in the warmer regions may enhance and even dominate
the formation of other O--bearing species, so that the determination of
its abundance is also  relevant
to understand the prevailing chemistry.

In this letter we present and analyze several OH far--IR lines,
involving levels  up to $\sim$420~K, toward the Starburst galaxy 
NGC~253 and the Seyfert 2 galaxy NGC~1068.
We compare these  extragalactic OH lines with our far--IR 
observations of Sgr~B2(M) and Orion~IRc2, and run simple non--LTE radiative
transfer models to qualitatively understand the behavior of the 
detected lines.

\begin{figure*} [ht] 
\centering
\includegraphics[angle=-90,width=16.5cm]{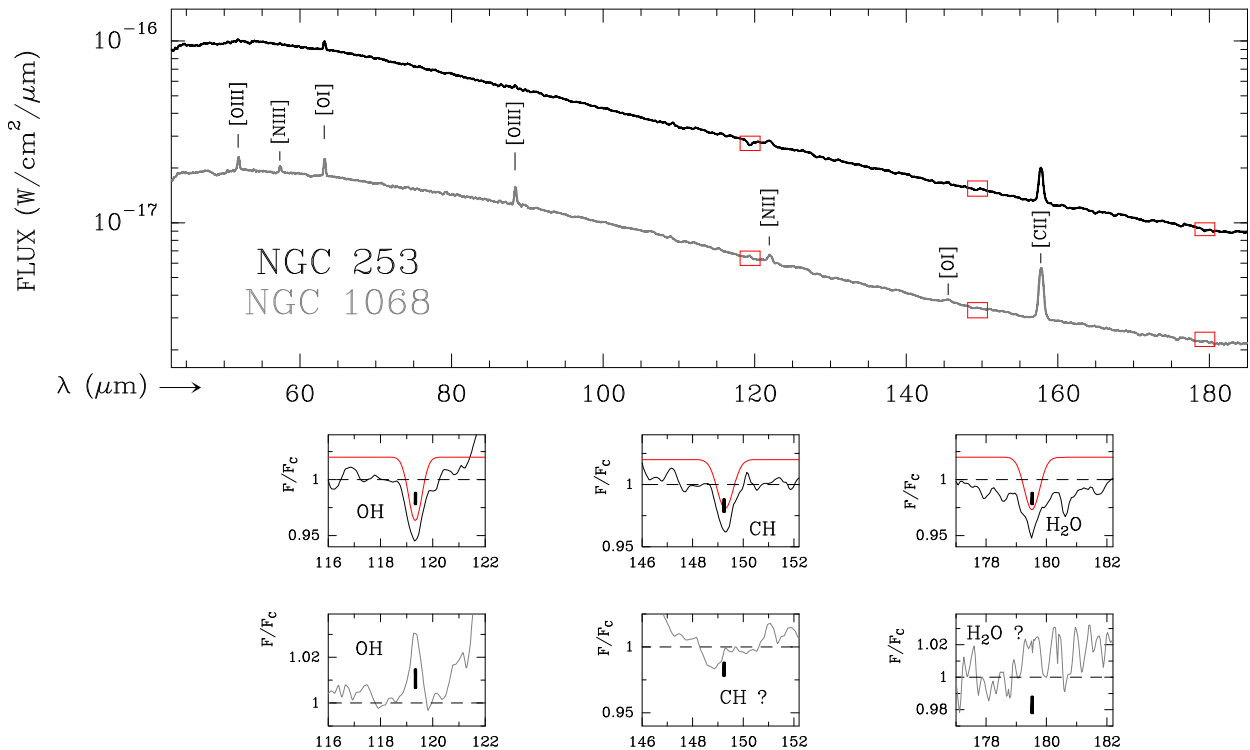}
\caption{ISO/LWS grating observations of NGC~253 and NGC~1068. The ordinate
corresponds to the flux and the abscissa is the
rest wavelength in $\mu$m. Atomic fine structure lines are labeled. Zooms
around selected wavelengths are also shown in the smaller panels with
models of the ground--state rotational lines of OH, CH and H$_2$O (see text). 
The ordinate of these panels is the continuum normalized flux.}
\label{gratings}
\end{figure*}

\section{Observations and Data reduction}

Several pure rotational lines of OH appear in the wavelength coverage of
the Long-- and Short--  Wavelength Spectrometers (LWS: Clegg et al. 1996; 
SWS: de Graauw et al. 1996) on board \textit{ISO} (Kessler et al. 1996).
Bradford et al. (1999) have previously presented 
LWS Fabry--Perot (FP) observations
of the OH ground--state line at $\sim$119~$\mu$m toward NGC~253 
($v_{LSR}\simeq250$~km~s$^{-1}$), 
while Spinoglio et al.  (1999) presented the full LWS grating spectrum of 
NGC~1068 ($v_{LSR}\simeq1140$~km~s$^{-1}$). In this work, we have used 
the public ISO Data  Archive (IDA) and analyzed the target dedicated time
numbers (TDTs) 24701103, 56901708 (for NGC~253) and 60500401 and 60501183
(for NGC~1068).
The spectral resolution of the LWS/grating is
$\lambda$/$\Delta\lambda$$\sim$200.
The OH $^2\Pi_{1/2}$--$^2\Pi_{3/2}$ $J$=5/2--3/2  cross--ladder line at 
$\sim$34~$\mu$m was observed with the SWS/grating with a spectral 
resolution of $\lambda$/$\Delta\lambda$$\sim$1500 (TDTs 37902123 and 64302927).
The LWS beam has a  $\sim$80$^{\prime\prime}$ diameter
while the SWS aperture in the $\sim$34 $\mu$m range is  
17$^{\prime\prime}$$\times$40$^{\prime\prime}$.
LWS and SWS  data were analyzed using the 
ISO Spectral Analysis Package\footnote{ISAP
is a joint development by the LWS and SWS Instruments Teams and Data Centers. 
Contributing institutes are CESR, IAS, IPAC, MPE, RAL, and SRON.}
(ISAP). Typical routines include: deglitching spikes due to cosmic rays,
oversampling and averaging individual scans, removing baseline polynomials and
extracting line fluxes by fitting gaussians.
More details can be found in the LWS handbook (Gry et al. 2003).
The full ISO/LWS spectra of  NGC~253 and NGC~1068 are shown
in Figure~1 while several OH rotational lines are shown in Figure~2.

\section{Results}

The far-IR spectrum of NGC~253 and NGC~1068 
(Fig.~1, \textit{large panel})
is dominated by the thermal emission of dust and by atomic fine structure 
emission lines (Negishi et al. 2001). The signal--to--noise ratio of these
spectra is lower than the \textit{ISO/LWS} spectrum of Arp~220, which
is very rich in molecular features (see Gonz\'alez--Alfonso et al. 2004).
However, the ground--state rotational lines of OH, CH and H$_2$O are 
clearly observed in NGC~253 and possibly in NGC~1068 
(Fig.~1, \textit{small panels}).
Figure~2 shows in more detail the OH $\sim$34, $\sim$53 and $\sim$79~$\mu$m
cross--ladder lines, the $\sim$84 and $\sim$119~$\mu$m lines of the
$^2\Pi_{3/2}$ ladder, and the $\sim$163~$\mu$m line of the $^2\Pi_{1/2}$ ladder
toward NGC~253 and NGC~1068.
For comparison we also show the OH spectra of archetype star forming regions
in our Galaxy: Sgr~B2 and Orion~IRc2. It is interesting to note that the behavior
of the OH lines  in NGC~1068 follows that of Orion~IRc2, i.e., emission in all the
observed lines (except in the $\sim$53  and $\sim$34~$\mu$m lines).
On the other hand, the OH lines in NGC~253 have a similar behavior than in
Sgr~B2(M), i.e., absorption in the $^2\Pi_{3/2}$ ladder and 
cross-ladder lines and emission in the $^2\Pi_{1/2}$ ladder
(Goicoechea \& Cernicharo 2002).
Hence, the OH rotational lines show distinct trends in two prototype galaxies 
where the nuclei activity is also different. 

Due to its large dipole moment and rotational constant, radiative pumping of
the rotational levels must be taken into account as long as collisions
to compute the OH level population.
For the typical temperatures of molecular clouds, LTE can only be 
achieved at very high H$_2$ densities, $n$(H$_2$)$>$10$^{10}$~cm$^{-3}$.
Hence, the knowledge of the dust continuum properties and its distribution
respect to the molecular gas is needed to analyze the OH excitation.

\begin{figure*} [ht] 
\centering
\includegraphics[angle=-90,width=17.5cm]{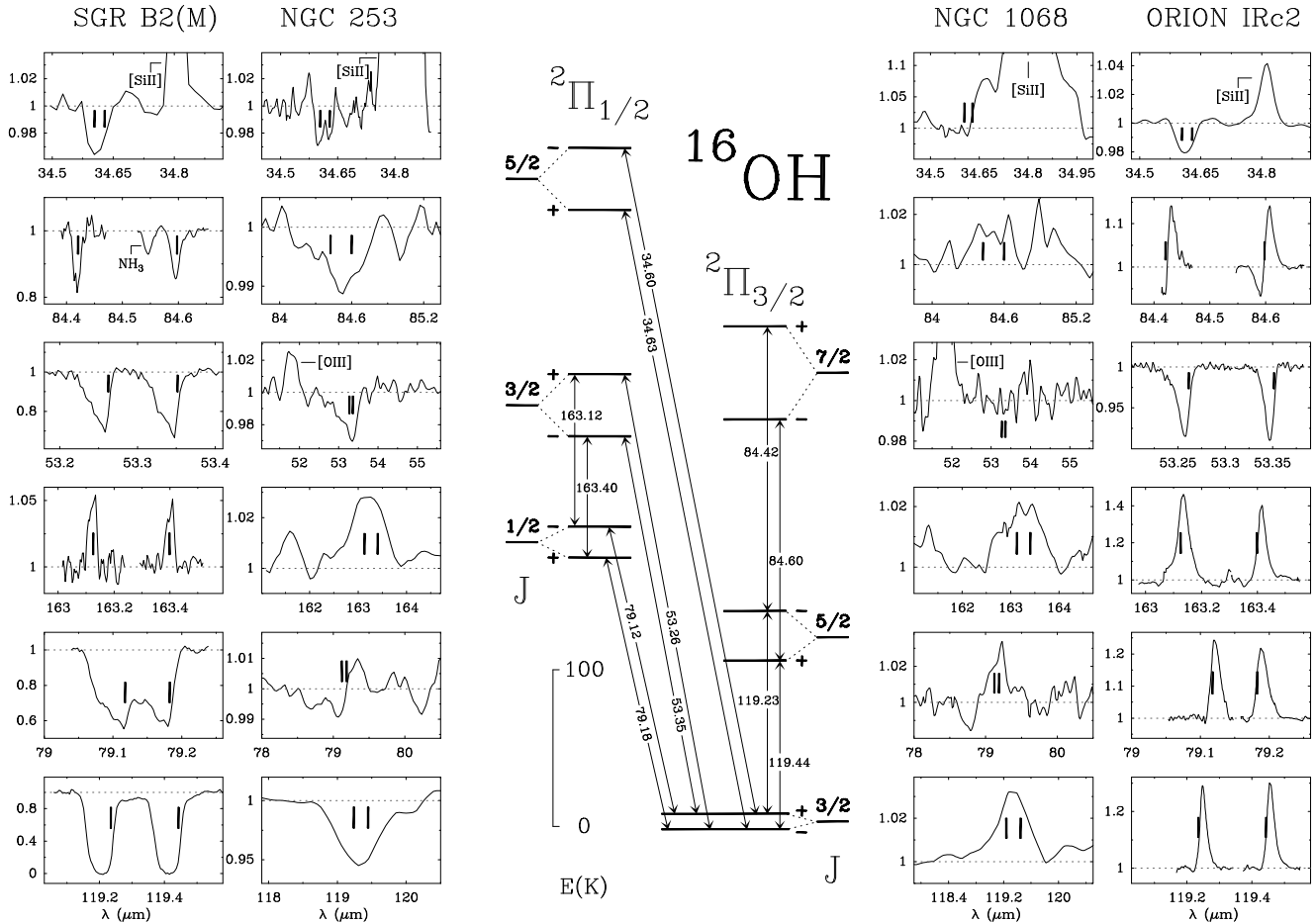}
\caption{ISO observations of OH rotational lines towards Sgr~B2(M) 
(from Goicoechea \& Cernicharo 2002), NGC~253, NGC~1068 and Orion IRc2. 
The ordinate scale corresponds to the continuum normalized flux. The
abscissa in all the boxes is the rest wavelength in microns. The vertical
thick lines indicate the rest wavelengths of the different OH $\Lambda$--doublets.
The $\Lambda$--doublets are spectrally resolved in the \textit{Fabry--Perot}
observations otherwise \textit{grating} observations are presented.
The rotational energy diagram of OH (neglecting hyperfine structure) and
the observed transitions are also shown in the middle of the figure.
The $\Lambda$--doubling has been exaggerated for clarity.} 
\label{observaciones}
\end{figure*}

\section{Discussion}

In order to qualitatively understand the contribution of the physical and 
geometrical parameters to the OH spectrum, we have modeled
the first 20 rotational levels of OH with two radiative transfer codes
(LVG and nonlocal; Cernicharo et al. 2000). 
The hyperfine structure of OH was not included in the calculations.
Collisional cross sections of OH with H$_2$
from Offer et al. (1994) have been used with an H$_2$ ortho--to--para
ratio of 3.

It is obviously more difficult to model the molecular emission/absorption
from a galactic nucleus (an ensemble of molecular clouds)
than to model a single molecular cloud. Since 
the observed extragalactic OH lines are not velocity resolved
and the angular resolution is larger than the dimensions
of the observed sources, we have just
considered a simplified model of an  \textit{IR core $+$ OH shell}
as a first approximation.
Therefore, we have  adopted an spherical geometry consisting of a shell
of radius R$_{shell}$ and continuum core 
of radius R$_{core}$. The exact radius 
determine the resulting OH column density and the properties
of the dust emission, however, \textit{ISO} observations
can not provide that information.
Then, we have compared the synthetic 
continuum--normalized spectrum with the \textit{ISO} observations considering that
the source is composed of several 
clouds with the same  physical conditions 
but do not overlap in the line--of--sight 
(see the discussion by Gonz\'alez--Alfonso et al. 2004).

In the case of NGC~253, the mid-IR continuum is confined to the
inner $\sim$10$''$ of the galaxy (Telesco et al. 1993).
We have taken this as a lower limit to R$_{core}$ in the far--IR.
In the other hand, low--$J$ CO  observations reveal that the molecular
gas extends up to $\sim$50$''$ (Harrison et al. 1999).
However, CO $J$=7--6 ($E_u$$\sim$150~K) observations show that 
the bulk of this emission is confined to the inner $\sim$15$''$ 
(Bradford et al. 2003). In addition, several OH masers at $\sim$18~cm
have been detected within the central $\sim$10$''$ of the nucleus
(Frayer et al. 1998).
We have assumed that the far--IR OH lines and the IR continuum
arises from the $\sim$15$''$ region, but we 
note that a more detailed model of NGC~253 could include an extended
component of low excitation OH gas (that will enhance the absorption
of the $\sim$119~$\mu$m fundamental line;
see Gonz\'alez--Alfonso et al. 2004 for the case of Arp~220).
In fact, observations of the extended absorption of OH 18~cm lines reveal
that OH in the ground--state is present as far as $\sim$45$''$ from
the nucleus (Turner 1985). 
However, here we are mainly interested in 
the general behavior of the higher energy OH lines
and since the ISO/LWS angular/spectral resolution is poor,
we will not add more complexity to the calculations in this work.
We have taken R$_{core}$/R$_{shell}$=0.9, although
we also investigate the effect of the dilution of the radiation
arising from the core. From a gray--body fitting of the LWS continuum
emission we have adopted  a 
dust opacity law $\tau_{\lambda}=\tau_{119}[119/\lambda(\mu m)]^{1.5}$
with $\tau_{119}$=1 and T$_d$=41~K. 

To compare with the NGC~1068 observations, we have considered that the OH emission
also arises from a single component with similar geometry and continuum emission
as for NGC~253.
However, our calculations showed that the OH emission fluxes could
only be reproduced if the emission arises from a more compact component with 
$\theta_{shell}$$\lesssim$10$''$ and  $\theta_{core}$$\sim$5$''$.
This suggests that the OH emission
may be dominated by dense gas in cloud cores near the nucleus.
This is the case of HCN observed at higher resolution 
(Jackson et al. 1993; Tacconi et al. 1994).
Note that OH masers at $\sim$18~cm have been found in the
inner arcsec of NGC~1068 (Gallimore et al. 1996).
Higher angular resolution is needed to
estimate the contribution of the Starburst ring to the OH emission.

\begin{figure} [h] 
\centering
\includegraphics[angle=0,width=8cm]{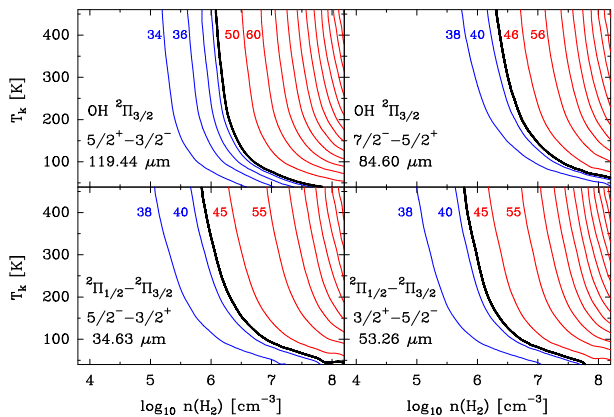}
\caption{Large Velocity Gradient models of different kinetic
temperature and density for the observed OH lines in NGC~253. The OH column
density is 6.5$\times$10$^{16}$~cm$^{-2}$. The thick contour corresponds to
the equivalent temperature of the continuum source 
(R$_{core}$/R$_{shell}$=0.9 and $\tau_{119}$(dust)=1).
To the left of this contour, lines are predicted in absorption.}
\label{fig3}
\end{figure}

Several results from the LVG computations are shown in Fig.~3. The excitation 
temperature of each rotational transition as a function of 
gas temperature (T$_k$) and density is
shown in each panel. The thick black contour represents the equivalent
temperature of the continuum. Therefore, to the left of this contour, the
OH lines are predicted in absorption. From the excitation models it
is clear that high densities are required to observe 
the OH lines in emission, i.e., the case of NGC~1068.
If  the R$_{core}$/R$_{shell}$ ratio is decreased
or the shell thickness is increased, the iso--T$_{ex}$ contours on Fig.~3
(the absorption/emission boundary),  will move to the left toward lower densities.
This is due to the increase of the reemission from the gas that do not 
absorb the core emission. 
Note that reemission also scales with the thickness of the shell.\\

To investigate the effect of the radiative coupling between components 
of different physical conditions we have also run several nonlocal models 
(see G\'onzalez--Alfonso \& Cernicharo 1993)
for OH including both the gas and dust transfer (Cernicharo et al. 2000). 
We run calculations for different geometrical and physical conditions.
As an example, Fig.~4 shows some of the results for the
$\sim$163.12 and $\sim$79.18~$\mu$m lines. 
The  $^2\Pi_{1/2}$ $J$=3/2--1/2 line at $\sim$163~$\mu$m  is
observed in emission in all the sources. 
Goicoechea \& Cernicharo (2002) showed that the $\sim$163~$\mu$m emission
in Sgr~B2(M), were the far--IR continuum is optically thick, is produced by a 
fluorescent--like mechanism in low density ($<$10$^4$~cm$^{-3}$) warm gas
(T$_k$$\simeq$150-300~K).
In the other hand, Melnick et al. (1987) showed that 
T$_k$$<$100~K and $n$(H$_2$)$>$10$^{6-7}$~cm$^{-3}$ are required to fit
the observed emission in Orion~IRc2 shocked region, so that collisions
dominate the excitation. 
However, these authors argued that even with those conditions, the radiative pumping
to the levels that give rise to the $\sim$163~$\mu$m emission has to be taken 
into account.

\begin{figure} [h] 
\centering
\includegraphics[angle=-90,width=8cm]{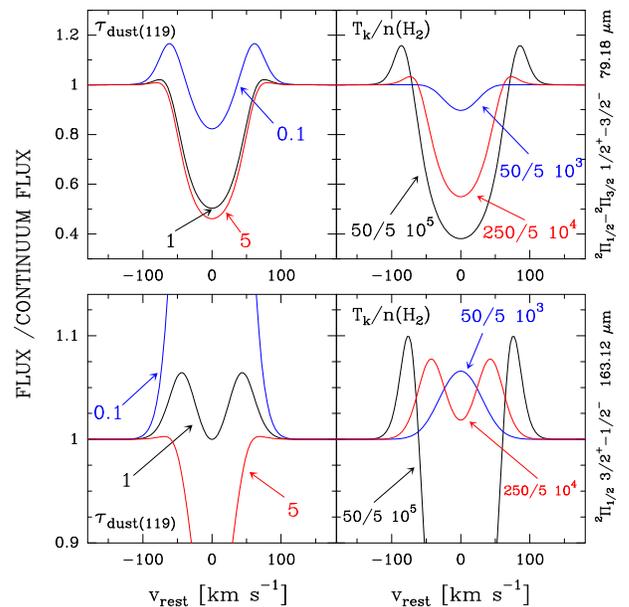}
\caption{Nonlocal models for the 
$^2\Pi_{1/2}$--$^2\Pi_{3/2}$ $J$=1/2$^+$--3/2$^-$ transition at
79.18~$\mu$m (\textit{upper panel}) and for the $^2\Pi_{1/2}$ $J$=3/2$^+$--2/2$^-$ 
transition at 163.12~$\mu$m (\textit{lower panel}). Line intensities 
have been normalized to the continuum emission. The abscissa is the rest
velocity. The velocity resolution in the line profiles is 1~km~s$^{-1}$.
All the models have $\chi$(OH)=10$^{-7}$.
\textit{Left panels}: Models with R$_{core}$/R$_{shell}$=0.9, T$_k$=150~K,
n(H$_2$)=5$\times$10$^4$~cm$^{-3}$ and variable dust opacity of the
continuum source.
\textit{Right panels}: Models with R$_{core}$/R$_{shell}$=0.9,
$\tau_{119}$(dust)=1 and variable T$_k$ and n(H$_2$).} 
\label{fig4}
\end{figure}

From our calculations it is clear that while at low OH abundances [$\chi$(OH)] the
$\sim$79, $\sim$53 and $\sim$34~$\mu$m cross-ladder absorption depths
are proportional to the OH column density [$N$(OH)], the $\sim$163~$\mu$m emission
over the continuum flux is nearly constant 
because the emission mechanism at low densities depends mainly on the 
continuum radiation.
At high $\chi$(OH), reemission start to  dominate the $\sim$79 and
$\sim$119~$\mu$m lines.
Taking into account the simplicity of our calculations, we find
$\chi$(OH)$\sim$10$^{-7}$ as the most realistic value for  NGC~253.
Lower values are required if the core emission is more diluted,
otherwise reemission dominates.
On the other hand, larger abundances can be possible 
if OH emission/absorption and  far--IR continuum emission
have different dilution factors.

Density and T$_k$ determine the role of collisional excitation. For low
$n$(H$_2$), radiative excitation dominates, and the OH spectrum is less
sensitive to $n$(H$_2$) and T$_k$ variations. This seems to be the case
of NGC~253 inner regions and the GC molecular clouds, which probably share
similar physical conditions. 
The effect of T$_k$ is more relevant 
for higher dilution of the core radiation. We found that the model with
T$_k$$\sim$150~K
and  $n$(H$_2$)$\lesssim$5$\times$10$^4$~cm$^{-3}$ 
better fits the OH observations of NGC~253. This temperature is in between
the T$_{ex}$ derived from the H$_2$ $S$(0) line by Rigopoulou et al. (2002)
and the T$_{k}$ derived from the CO $J$=7--6 line emission
by Bradford et al. (2003). 
Finally, from the ground--state rotational lines of CH and H$_2$O 
toward NGC~253 and assuming intrinsic line widths of $\sim$200 km~s$^{-1}$ and
T$_{ex}$$<$T$_d$ (absorption lines) we found a lower limit for the
CH and H$_2$O column densities of $\sim$5$\times$10$^{13}$ and 
$\sim$1$\times$10$^{15}$~cm$^{-2}$ respectively (Fig.~1).

Similar arguments also apply for NGC~1068.
Following Sternberg et al. (1994) we have first assumed T$_k$$=$50~K.
Taking into account the simplifications in our models, 
the calculations for NGC~1068 show that the density must be 
larger than 10$^6$~cm$^{-3}$, otherwise the OH lines are still predicted 
in absorption. 
For $n$(H$_2$)=5$\times$10$^5$~cm$^{-3}$, the temperature
has to be increased to $\sim$150~K to fit the observations, so that it is
likely that a fraction of warm neutral gas also exists in the
clouds near the nucleus of NGC~1068.
Intense emission in other OH  rotational lines (not detected 
by \textit{ISO}) are predicted for higher densities and
temperatures.\\

\section{Conclusions}

Irrespective of the specific nuclei scenario and of the density of
the dominant molecular clouds, OH observations show that
a similar component of warm neutral gas (T$_k$$\sim$150~K)
exists in NGC~253 and NGC~1068.
This component can influence the chemistry as neutral--neutral reactions
involving OH start to be efficient for the production of 
other oxygen molecules such as H$_2$O and SiO.
The contribution of PDRs, shocks, X--rays and cosmic
rays to the heating of such component
in Seyfert vs. Starburst galaxies is still far from settled. 
Future works using diagnostics
of the warm gas will be extremely useful.
In particular, several rotational lines of light hydrides 
such as OH, CH and H$_2$O will be observed 
at larger angular resolution by the
HIFI and PACS instruments, on board Herschel
in many galactic and extragalactic sources.

\acknowledgments

We thank Spanish DGES and PNIE for funding
support under grants PANAYA2000-1784, ESP2001-4516, AYA2002-10113,
ESP2002-01627 and AYA2003-02785.


\begin{thebibliography}{}


\bibitem[]{Bet89}
Betz, A.L., \& Boreiko, R.T. 1989, ApJ, 346, L101

\bibitem[]{Bra99}
Bradford, C.M.,  et al. 1999,
The Universe as Seen by ISO. Eds. P. Cox \& M. F. Kessler. ESA-SP 427., p. 861

\bibitem[]{Bra03}
Bradford, C.M.,  et al. 2003, ApJ, 586, 891


\bibitem[]{Cer00}
Cernicharo, J., Goicoechea, J.R. \& Caux, E.
2000, ApJ, 534, L199


\bibitem[]{Cle96}
Clegg, P.E., et al. 1996, A\&A, 315, L38

\bibitem[]{Gra96}
de Graauw, T. et al. 1996, 315, L49 

\bibitem[]{Fra98}
Frayer, D. T., Seaquist, E. R., \& Frail, D. A. 
1998, AJ, 115, 559

\bibitem[]{Gal96}	
Gallimore, J. F., Baum, S. A., O'Dea, C. P, Brinks, E., \& Pedlar, A.
1996, ApJ, 462, 740

\bibitem[]{Goi02}
Goicoechea, J.R., \& Cernicharo, J.
2002, ApJ, 576, L77


\bibitem[]{Goi04}
Goicoechea, J.R., Rodr\'{\i}guez-Fernandez, N.J. \&  Cernicharo, J.
2004, ApJ, 600, 214

\bibitem[]{Gon93}
G\'onzalez--Alfonso, E. \&  Cernicharo, J.
1993, A\&A, 279, 503

\bibitem[]{Gon04}
G\'onzalez--Alfonso, E., Smith, H,A., Fisher, J. \& Cernicharo, J.
2004, ApJ, 613, 247

\bibitem[]{Gry03}
Gry, C. et al. 2003, 
The ISO Handbook', Volume III - LWS - The Long Wavelength Spectrometer Version 2.1.
ESA SP-1262, 2003.


\bibitem[]{Harr99}
Harrison, A.,  Henkel, C., \& Russell, A.
1999, MNRAS, 303, 157

\bibitem[]{Jack93}
Jackson, J.M., Paglione, T.A. D., Ishizuki, S., \& Nguyen-Q-Rieu
1993, ApJ, 418, L13


\bibitem[]{Kes96}
Kessler, M.F., et al. 1996, A\&A, 315, L27

\bibitem[]{Mel87}
Melnick, G.J., Genzel, R., \& Lugten, J.B. 1987, ApJ, 321, 530

\bibitem[]{Mel90}
Melnick, G.J., Stacey, G.J., Genzel, R., Lugten, J.B. \& Poglitsch, A.
1990, ApJ, 348, 161

\bibitem[]{Neg01}
Negishi, T., Onaka, T., Chan, K.-W., \& Roellig, T. L.
2001, A\&A, 375, 566


\bibitem[]{Off94}
Offer, A.R., van Hemert, M.C., \& van Dishoeck, E.F. 1994, JChPh, 100, 362  

\bibitem[]{Ran95}
Randell, J., Field, D., Jones, K. N., Yates, J. A. \& Gray, M. D.
1995, A\&A, 300, 659

\bibitem[]{Ri02}
Rigopoulou, D., Kunze, D., Lutz, D., Genzel, R. \&  Moorwood, A. F. M.
2002, 389, 374

\bibitem[]{Rod01}
Rodr\'{\i}guez--Fernandez, N.J. et al. 2001, A\&A, 365, 174

\bibitem[]{Ski97}
Skinner, C. J., Smith, H. A., Sturm, E., Barlow, M. J., Cohen, R. J. \& Stacey, G. J.
1997, Nature, 386, 472

\bibitem[]{Spi99}
Spinoglio, L. et al. 1999.
The Universe as Seen by ISO. Eds. P. Cox \& M. F. Kessler. ESA-SP 427., p. 969


\bibitem[]{Ste94}
Sternberg, A., Genzel, R., \& Tacconi, L. 1994,
ApJ, 436,  L131

\bibitem[]{Tac94}
Tacconi, L. J., Genzel, R., Blietz, M., Cameron, M., Harris, A. I., \& Madden, S.
1994, ApJ, 426, L77

\bibitem[]{Sto81}
Storey, J., Watson, D.M., \& Townes, C.H. 1981, ApJ, 244, L27

\bibitem[]{Tel93}
Telesco, C. M., Dressel, L. L., \& Wolstencroft, R. D., 1993,
ApJ, 414, 120.

\bibitem[]{Tur85}	
Turner, B. E. 1985, ApJ, 299, 312

\bibitem[]{Vi85}
Vicuso, P.J., et al. 1985, ApJ, 296, 149

\bibitem[]{Wat85}
Watson, D.M., Genzel, R., Townes, C.H., \& Storey, J.W.V. 
1985, ApJ, 298, 316

\end{thebibliography}
\end{document}